\documentclass[aps,showpacs,floatfix]{revtex4}
\usepackage{graphics}
\usepackage{dcolumn}
\newcommand{\bea}{\begin{eqnarray}}
\newcommand{\eea}{\end{eqnarray}}
\newcommand{\be}{\begin{equation}}
\newcommand{\ee}{\end{equation}}

\def\be{\begin{eqnarray}}
\def\ee{\end{eqnarray}}
\def\bd{\begin{displaymath}}
\def\ed{\end{displaymath}}

\def\etal{{\em et al.}}

\def\NP{Nucl. Phys. }
\def\PR{Phys. Rev. }
\def\PRL{Phys. Rev. Lett. }
\def\PL{Phys. Lett. }

\begin{document}
\title
{Role of pairing interaction in neutron rich odd and even Zr nuclei }
\author{Madhubrata Bhattacharya}
\author{G. Gangopadhyay}
\email{ggphy@caluniv.ac.in}
\affiliation{Department of Physics, University of Calcutta\\
92, Acharya Prafulla Chandra Road, Kolkata-700 009, India}

\begin{abstract}
Neutron rich Zr nuclei with number of neutrons between $N$= 50 and 82
are investigated in the relativistic mean field approach in co-ordinate
space. The resonant levels in the positive energy continuum have been
explicitly included in the calculation. Odd nuclei have been treated in the 
blocking approximation. Our calculation indicates that the dripline for 
odd mass isotopes is far away from that for the even mass ones.
Pairing 
interaction plays a significant role in stabilizing the even isotopes, thus 
extending the dripline for them. 
\end{abstract}
\maketitle

Exotic nuclei  very close to the neutron drip line
exhibit a number of new and interesting features such as neutron skin and halo, 
rearrangement of magic numbers, quenching of spin orbit splitting, etc. The 
last bound neutrons in such a nucleus lie very close to the continuum. Hence, 
the effect of the positive energy continuum on the structure of such nuclei 
needs to be studied carefully. Another very important aspect of 
nuclei near the drip line is the additional stability contributed by the 
pairing. This effect may be studied
by investigating the odd mass nuclei along with the even mass ones. 
Although a number of such investigations has been undertaken in lighter mass 
regions, nuclei in the the medium and the heavy mass regions, have not yet been 
examined  in sufficient detail. 

In the present work, we study  Zr 
isotopes using relativistic mean field (RMF) approach in co-ordinate space.
Though the RMF equations are usually solved by expanding 
them in a harmonic oscillator basis, it is well known that this
method cannot explain the density in halo 
nuclei near the drip line because of slow convergence. Relativistic Hartree 
Bogoliubov method in co-ordinate space (RCHB) has been found to be a very
accurate method of treating the nuclei very close to the drip line. The 
effect of the 
states in the continuum has been incorporated in most of these calculations by 
solving the equations in co-ordinate space where the continuum has been replaced
by a set of discrete positive energy states
with the box normalization condition.
However, the single particle energy levels depend on the size of 
the box chosen in this scenario, an unwelcome complication.

Nonrelativistic mean field equations involving continuum states have already 
been
solved with exact boundary conditions\cite{hfb1,hfb2} for zero range and
finite range pairing forces.
Corresponding relativistic calculations have also been 
performed\cite{rmfc,rmfc1}
with exact boundary conditions. All these calculations
have taken into account the width of the continuum levels also. 
We have also used the RMF calculation in co-ordinate space
including the continuum states to study neutron rich C and Be\cite{C} and 
Ca and Ni nuclei\cite{CaNi}. We find that this method gives results in very 
good agreement 
with the more complicated relativistic Hartree Bogoliubov (RHB) approach. 

In the present work, the structure of even and odd Zr isotopes with neutron 
number between $N=50$ and 82 have been studied in the 
relativistic mean field formalism with exact boundary conditions, 
Numerous authors have studied neutron rich even mass Zr
isotopes up to the neutron drip line using relativistic and
nonrelativistic mean field approaches. We cite only a few works
published recently. Sharma \etal\cite{Sh} first applied RMF to study 
neutron rich Zr nuclei up to the drip line. RHB approach was also utilized to 
study some of the neutron rich isotopes\cite{Zr1,Zr2}. Zhang \etal \cite{Zr3}
used RCHB approach to study Zr and some other proton closed shell nuclei.  
Deformed RMF approach was used by Geng \etal ~to study even-even Zr 
nuclei\cite{Geng}. RCHB approach was applied to study $^{134}$Zr\cite{Meng}.
Pearson\cite{Pearson} compared the results of Skyrme HF and RMF calculations for
even-even neutron rich Zr nuclei. 
Recently co-ordinate space Hartree-Fock-Bogoliubov calculations were performed
by Blazkiewicz \etal\cite{hfb} for even-even Zr isotopes. They found that the 
neutron drip line is at $N=82$. 
Continuum RMF calculation with exact boundary conditions has been used to 
study Zr isotopes\cite{rmfc}. Zhang \etal\cite{Zh} have studied $^{122}$Zr in 
the continuum relativistic mean field theory using the analytic continuation 
approach for single-particle resonance. The authors of the last two
works have used the relativistic force NLSH\cite{NLSH} though Sandulescu \etal
~have concluded that the results are not much different even if one uses 
NL3\cite{NL3} or
TM1\cite{TM1}. They have studied the even-even nuclei beyond N=82. However, we have come 
across no mean field calculation in neutron rich odd mass Zr nuclei. 

As already mentioned, the aim of the present work is to study the role of the 
pairing  as well as the resonant states in neutron rich Zr nuclei, particularly 
with respect to odd mass nuclei near the neutron drip line. We have earlier 
seen\cite{CaNi} that in the case of Ca isotopes, the drip line for the even 
mass nuclei lies far beyond the same for odd mass ones. However, for Ni 
isotopes, the 
dripline nuclei for both even and odd masses  are adjacent. This phenomenon has 
been interpreted as the effect of pairing correlations stabilizing the
even-even nuclei in the case of levels with appreciable occupation factor 
lying very close to or even inside the positive energy continuum. This 
scenario is observed in the single particle energy levels in Ca but not in Ni
isotopes. In the present work we extend our study to odd mass Zr nuclei. 

Relativistic mean field theory is well known and will not be detailed here.
In the present work, we have mainly employed the force NL3\cite{NL3}. 
This force is known to provide good results for binding energy and 
radius throughout the periodic table. For comparison, we have also redone 
the calculations with the force NLSH\cite{NLSH}.
 Spherical symmetry has been assumed for all the nuclei as the protons form a 
closed shell. It is likely that some nuclei in the chain 
are deformed and also possible that the actual position of the neutron 
drip line may vary slightly on inclusion of deformation. However, our aim is 
to investigate a possible 
difference  in the location of the drip line in even and odd mass nuclei.
Our conclusions in this regard, we believe, are more robust and not likely 
to be affected by inclusion of deformation to any large extent.

A delta interaction has been used for the pairing correlations  
between neutrons, {\em i.e.}
$V=V_0\delta(\vec{r}_1-\vec{r}_2)$. The usual BCS equations now contain
contributions from the bound states as well as the resonant continuum. The
equations involving both these types of states have already been obtained
\cite{new1,new2} and have been referred to as resonant-BCS (rBCS) equations.
We have included the effect of the width of the positive energy levels.
These equations have been solved in the co-ordinate space on a
grid of size 0.08 fm. The positive energy resonance solutions are
obtained using the scattering approach. All the negative energy states beyond
N=40 as well as the positive energy states  for which resonance solutions have
been found have been included in the rBCS calculation.
We have assumed that beyond 20 fm, the effect of nuclear 
interaction vanishes. We choose $V_0$= -575 MeV for the strength of the 
delta-interaction as this gives very good results for binding energy of
even even Zr nuclei close to the stability valley.
Odd nuclei have been treated in the blocking approximation corresponding to
different single particle states.

The ground state spin-parity of the odd isotopes are known up to $N=63$,
{\em i.e.} $^{103}$Zr. In our calculation for odd isotopes, 
the ground state spin parity values come out to be $5/2^+$ up to $^{97}$Zr,
$1/2^+$ for nuclei between $^{99}$Zr and $^{111}$Zr and $11/2^+$ beyond that.
We find that  our calculations can correctly explain
the ground state spin-parity for nuclei up to $^{95}$Zr and $^{99}$Zr. 
In $^{97}$Zr, the ground state has actual spin parity $1/2^+$. Our calculation 
predicts the ground state to be $5/2^+$ though there is another state with
spin parity $1/2^+$ approximately 270 keV above this state. Similarly, the 
ground state in $^{101}$Zr is predicted to be $1/2^+$. The spin parity
of the actual ground state is $3/2^+$. The corresponding theoretical state 
is at an excitation energy of 530 keV. Our model fails to explain the 
observed negative parity of the ground state of $^{103}$Zr which may possibly 
be due to deformation. Finally, we should mention that there are some doubts to 
the experimental spin-parity assignment to the ground states of 
$^{99,101,103}$Zr.

To check whether our results are dependent on the force selected,
all the calculations have been redone using the  force NLSH. Even Zr nuclei 
have already been studied using this force in the present model by 
Sandulescu \etal\cite{rmfc}. Following them we have taken $V_0=-275$ MeV.
In all the odd isotopes, the results for the ground state spin parity remain
unchanged. Particularly in $^{97}$Zr, our calculation gives the ground state 
to be $5/2^+$ where it is about 290 keV below the  $s_{1/2}$ state.

The preceding results show that the the ground state spin is correctly 
reproduced in lighter nuclei. In Table \ref{1q} we present some of our results 
for one-quasiparticle states in the lighter nuclei where 
some information is available in the low energy regime, i.e. $^{91,93,95}$Zr. 
The experimental values are for the one-quasiparticle levels built on the 
corresponding states
given in table. For example, in $^{93}$Zr a $3/2^+$ state has been observed
at 0.26 MeV by (d,p) reaction. However, study \cite{Zr4} suggests that 
it is in reality a three-quasiparticle state with $(2d_{3/2})^3_{3/2^+}$
configuration and the one-quasiparticle state actually corresponds to the
$3/2^+$ level at  1.4 MeV. It is seen that the above results for ground state 
spin parity values as well as for the other low energy one-quasiparticle states 
do not vary substantially if the force is
changed from NL3 to NLSH. Henceforth, all the results to be presented are for
the force NL3 unless otherwise mentioned. We also observe that the experimental
values are reproduced reasonably well in most cases. 
One has to remember that in odd nuclei, often there is some contribution from 
the three-quasiparticle 
configurations even to low lying states and thus the experimental results may 
not agree with mean field theory without taking this configuration mixing into 
account.

\begin{figure}[h]
\resizebox{4in}{!}{ \includegraphics{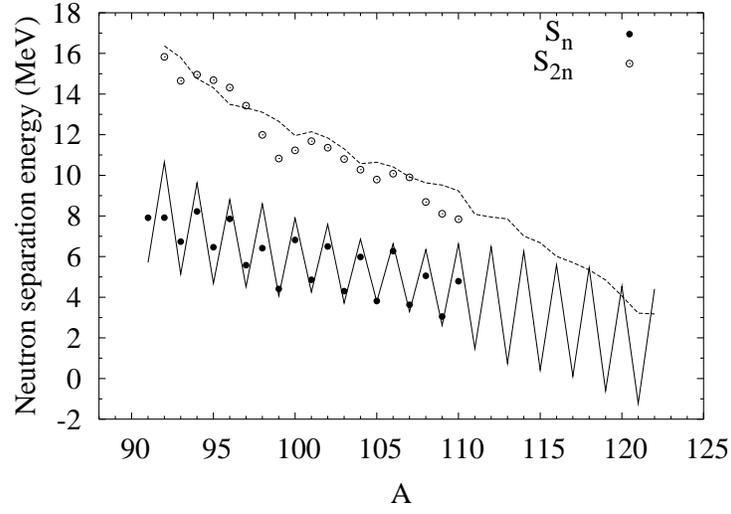}}
\caption{Calculated and experimental one and two neutron separation energy 
in Zr isotopes. Theoretical results for one (two) neutron separation energies
are connected by solid (dashed) lines  while experimental values are indicated
by hollow (filled) circles.\label{nfig}}
\end{figure}

In Table \ref{be}, the calculated and experimental or estimated binding energy 
values for the odd and even Zr isotopes between $N=50$ and 82  are 
presented. For comparison,
the results for even-even nuclei from RCHB approach of Zhang \etal\cite{Zr3}
using the force NLSH are also given.
One can see the excellent overall
agreement between experiment and theory particularly for the even isotopes
in our calculation. 
In Fig. \ref{nfig}, we plot the one and two neutron separation energy for the 
nuclei studied. One neutron separation energy for a nucleus with neutron number
$N$ has been written as
$S_n={\rm B.E.}(Z,N)-{\rm B.E.}(Z,N-1)$,
and similarly for two neutron separation energy $S_{2n}$. It is true that 
some of the $S_n$ values are large compared to the experiment. 
Besides, recent high precision mass measurements \cite{mass1}
show interesting structures beyond $N=56$.
However, nuclear
structure calculations in this regard are often not exact and do not agree
with experiment. Even more sophisticated models like RHB can not predict the
results exactly. We should remember that the effects of deformation and 
configuration mixing have not been included
which may affect the results. The most interesting aspect of the above results 
is the dripline for odd isotopes. 

It is well known that relativistic theories predict that the 
dripline for even Zr nuclei is around $N=96$. We have confirmed that our 
calculations agree with them. In the case of odd isotopes, we find that the 
dripline is at $N=77$ for the force NL3. It is true that the location of the
 drip line 
is not exactly fixed in view of the assumptions in calculation, {\em viz.} 
neglecting the deformation, disregarding the effect of three-quasiparticle 
configurations, etc. However, the general result on the position of
the drip line are expected to remain unaltered. Also interesting is the fact 
that the force NLSH predicts the dripline for odd nuclei to be at $N=81$. 
This is due to the use of a pairing strength value only half as large as that 
of 
the NL3 calculation. As explained later, because of the larger pairing 
strength, the even mass nuclei are more bound in the latter case and 
consequently the dripline for odd nuclei is at a lower mass. 
Though the two results do not agree 
exactly, we see that out essential conclusion that the driplines for the even 
and odd mass isotopes are far from each other, remains unaltered. In our 
earlier study involving odd mass Ca and Ni isotopes, we found that the dripline 
in odd nucleus is sensitively dependent 
on the pairing interaction as well as the single particle levels in neutron 
rich nuclei. Any reordering in the single particle level structure may 
introduce a corresponding change in the magic number in that mass region. The 
pairing interaction also plays an important role stabilizing the even-even 
isotopes compared to the neighbouring odd mass ones. 

\begin{figure}
\resizebox{4in}{!}{ \includegraphics{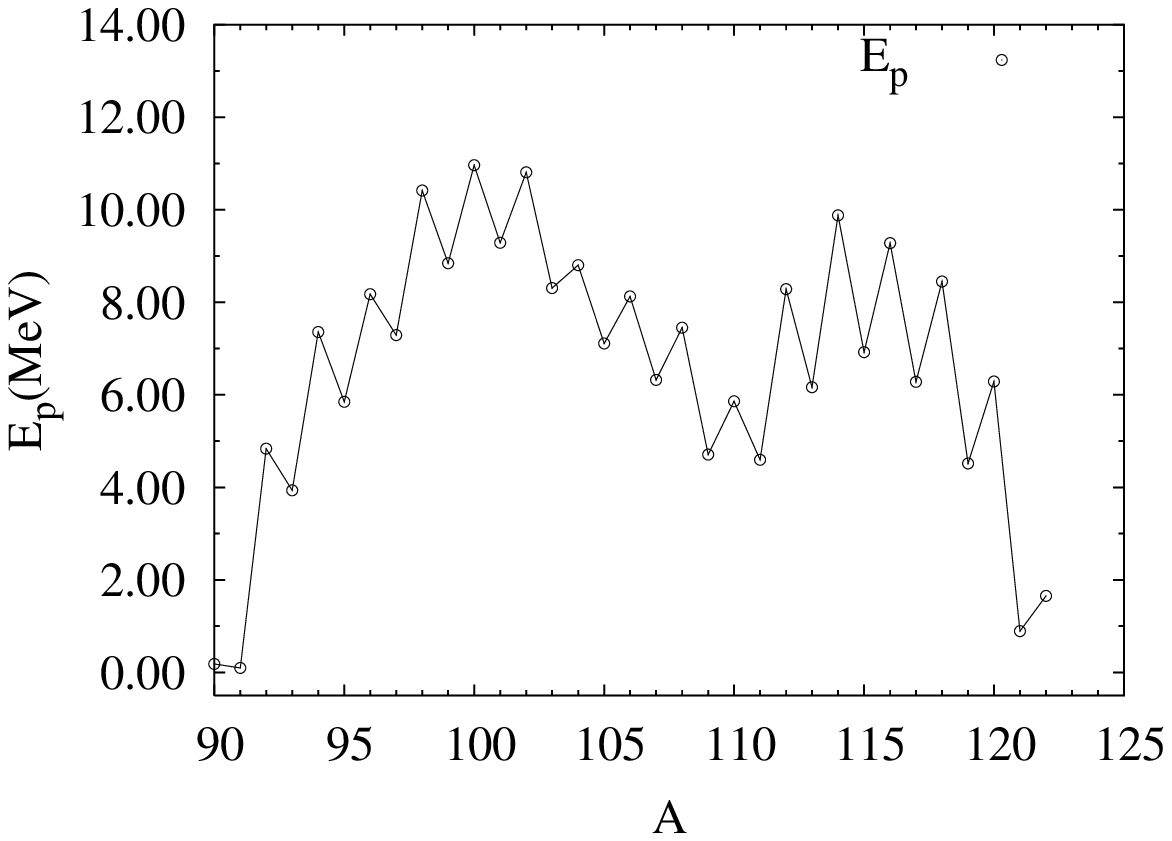}}
\caption{Correlation energy in  
in Zr isotopes\label{pairfig}}
\end{figure}

We plot the correlation energy due to pairing obtained in our calculation in 
Fig. \ref{pairfig}.
The correlation energy $E_p$ has been defined as the difference between the 
binding energy values the cases where pairing interaction has been taken 
into account and where it has been neglected. In odd nuclei, the ground state
spin-parity for the solution when the pairing has been switched off may not 
agree with that when pairing is present. For example, in $^{111}$Zr, the 
extreme single particle model suggests that the last neutron should be in 
the $1h_{11/2}$ state with the $3s_{1/2}$ being completely filled. However, 
calculations predict that this corresponds to actually an excited state. The 
ground state has a neutron pair in the $1h_{11/2}$ state and the odd neutron 
in the $3s_{1/2}$ state. The correlation energy for a pair of neutron in the high 
spin state is very large, thus pushing down the latter configuration to be the 
ground state. The odd-even staggering is clearly visible, particularly in the case of 
nuclei with $N\ge 113$, where the last odd neutron occupies the $1h_{11/2}$ 
single particle orbit in the ground state. This 
large difference in the pairing energies in even and odd isotopes can explain 
the additional stability of the even-even isotopes. In nuclei with $N\ge 113$,
the correlation energy for odd and even isotopes differ by more than 3 MeV. This
is sufficient to bind the even-even isotopes though the odd isotopes beyond
$N=117$ are unstable against one neutron emission. 

\begin{figure}
\resizebox{4in}{!}{ \includegraphics{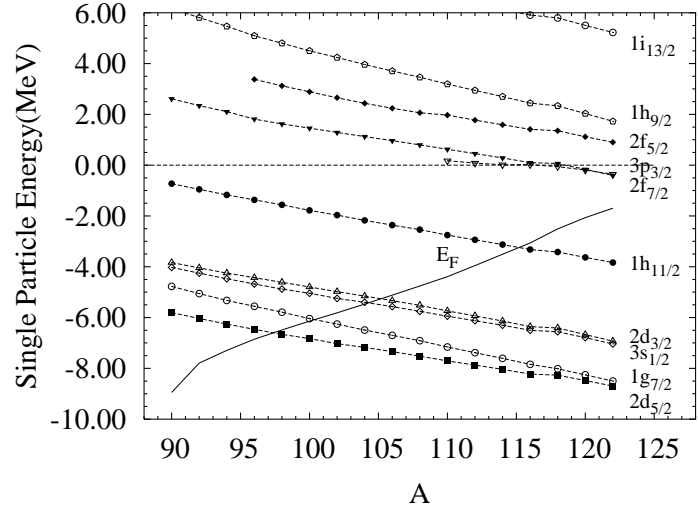}}
\caption{Single particle levels in even  Zr isotopes\label{levfig}}
\end{figure}

One interesting observation emerging from correlation energy results  
is the reduction in  the magic gap at $N=82$. The correlation energy 
does not vanish but has a value of 1.65 MeV. One also sees that the
correlation energy decreases at $N=70$. In Fig. \ref{levfig}, we plot the single 
particle levels near the Fermi level for the even isotopes. One can see that 
the large gap observed at $N=82$ in nuclei near the stability valley is not 
present. Actually, in neutron rich nuclei, the spin orbit splitting is reduced 
and the $1h_{11/2}$ level is higher in the Zr nuclei studied. So we observe 
smaller gaps at $N=70$ and $N=82$ rather than a single large gap at the 
conventionally known magic number $N=82$. Such a rearrangement of the single
particle orbitals is also observed for the NLSH results.

\begin{figure}
\resizebox{4in}{!}{ \includegraphics{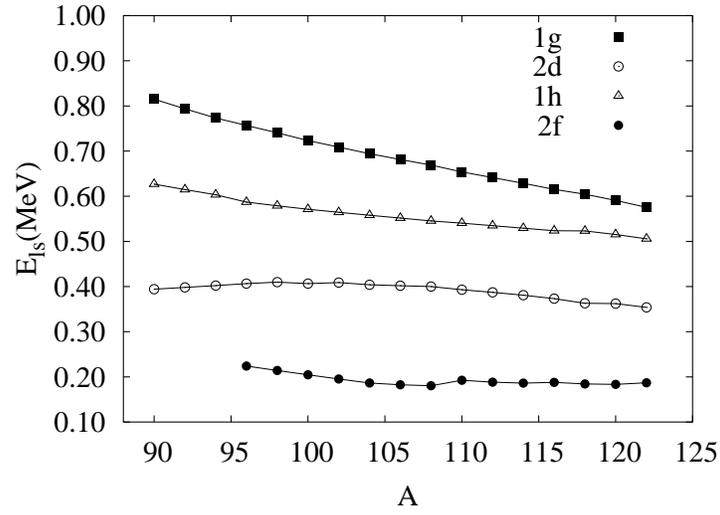}}
\caption{Spin orbit splitting in even 
 Zr isotopes\label{so}}
\end{figure}

One of the interesting predictions in neutron 
rich nuclei is this decrease in the spin-orbit splitting. Indeed it is 
particularly this effect that leads to the prediction and, in light nuclei, 
observation of emergence of 
new magic numbers along with disappearance of old ones. Though we have not 
studied the drip line for even even isotopes which, as we have observed,
is far away form the isotopes we have studied, a decrease in the spin
orbit splitting is already evident. In Fig. \ref{so}, we have plotted the 
splitting for these nuclei. Here we have defined 
$E_{ls}=|E_{l+1/2}-E_{l-1/2}|/(2l+1)$. It is clear that for high spin 
orbitals, the 
splitting falls off sharply with increase in the number of neutrons.

Neutron rich Zr nuclei with number of neutrons between $N$= 50 and 82
are investigated in the relativistic mean field calculation in co-ordinate
space. The resonance levels in the positive energy continuum have been
explicitly included in the calculation by using the scattering approach. 
Odd nuclei have been treated in the blocking approximation 
A zero range 
force has been used for interaction between neutrons. Our calculation indicates 
that the odd mass neutron dripline Zr nucleus is $^{117}$Zr. It is far away 
from the even mass dripline nucleus $^{136}$Zr predicted by earlier studies. 
The pairing interaction plays a significant role in stabilizing the even 
isotopes. The spin orbit splittings for the high spin single particle
orbitals decrease drastically for large neutron excess, thus leading to 
a reduction in the magic gap an $N=82$.

This work was carried out with financial assistance of the
Board of Research in Nuclear Sciences, Department of Atomic Energy (Sanction
No. 2005/37/7/BRNS), Government of India.

\clearpage

\begin{table}[h]
\begin{center}
\caption{Excitation energies (in MeV) of one-quasiparticle states in 
$^{91,93,95}$Zr. State refers to the single particle state on which the
one-quasiparticle state is built.
\label{1q}}
\begin{tabular}{ccccc}\hline
A &State &  NL3 & NLSH & Expt.\\\hline
   & $2d_{3/2}$ & 1.9 & 1.9 & 2.0\\
91 & $3s_{1/2}$ & 1.7 & 1.6 & 1.2\\
   & $1g_{7/2}$ & 1.1 & 1.1 & 1.8\\\hline
   & $2d_{3/2}$ & 1.5 & 1.5 & 1.4\\
93 & $3s_{1/2}$ & 1.2 & 1.2 & 0.9\\
   & $1g_{7/2}$ & 1.1 & 1.0 & 1.4\\\hline
   & $2d_{3/2}$ & 1.1 & 1.2 & 1.1\\
95 & $3s_{1/2}$ & 0.6 & 0.8 & 0.9\\
   & $1g_{7/2}$ & 0.9 & 1.0 & 1.6\\\hline
\end{tabular}
\end{center}
\end{table}

\begin{table}[h]
\begin{center}
\caption{Binding energy values in Zr isotopes. The values are in MeV.
Here Expt. refers to either experimental or estimated values and are from
\cite{mass} except for the nuclei with A=98-105 where they are from
\cite{mass1}. The RMF column indicates the theoretical results obtained in the 
present work while RCHB refers to the Relativistic continuum
Hartree Bogoliubov results of Zhang \etal
\cite{Zr3}.\label{be}}
\begin{tabular}{cccc|cccc}\hline
A&Expt. & RMF &RCHB&A&Expt. & RMF&RCHB \\\hline
90&783.892&783.432&783.173&
107&887.565&889.238\\
91&791.806&789.145&&
108&892.620&895.580&883.105\\
92&799.721&799.802&795.603&
109&895.672&898.198\\
93&806.456&804.940&& 
110&900.460&904.816&891.983\\
94&814.677&814.566&807.665& 
111&&906.286\\
95&821.139&819.248&&
112&&912.766&900.136\\
96&828.995&828.058&819.421&
113&&913.513\\
97&834.571&832.570&&
114&&919.775&907.646\\
98&840.986&841.163&830.903&
115&&920.192\\
99&845.394&845.227&&
116&&925.786&914.714  \\
100&852.214&853.117&842.122& 
117&&925.900\\
101&857.073&857.372&&
118&&931.356&921.459     \\
102&863.571&864.960&853.053&
119&&930.747 \\
103&867.871&868.673&&
120&&935.292&927.963     \\
104&873.851&875.534&863.613&
121&&934.062 \\
105&877.664&879.316&&
122&&938.468&934.285    \\
106&883.934&885.949&873.647\\
\hline
\end{tabular}
\end{center}
\end{table}
\clearpage

\end{document}